# Infrared metamaterials by RF sputtered ZnO/Al:ZnO multilayers

Kevin C. Santiago,[1*] Rajeh Mundle,[2] Curtis White,[3] Messaoud Bahoura,[4] and Aswini K. Pradhan[5*]

*Electrical and Computer Engineering Department, Tennessee State University, Nashville, TN 37209, United States*

*Center of Materials Research, Norfolk State University, Norfolk, VA 23504, United States*

Hyperbolic metamaterials create artificial optical anisotropy using metallic wires suspended in a dielectric medium, or alternating layers of a metal and dielectric. Here we fabricated ZnO/Al:ZnO (AZO) multilayers by the RF magnetron sputtering deposition technique. Our fabricated multilayers satisfy the conditions required for a type I hyperbolic metamaterial. The optical response of individual AZO and ZnO films, as well as the multilayered film were investigated via angular transmittance and spectroscopic ellipsometry measurements. The optical response of the multilayered system is predicted using the nonlocal-corrected Effective Medium Approximation (EMA). The spectroscopic ellipsometry data of the multilayered system was modeled using a uniaxial material model and EMA model. Both theoretical and experimental studies validate the multilayers undergo a hyperbolic transition at a wavelength of 2.2 µm.

**Introduction**

Metamaterials (MMs) are man-made materials that exhibit properties nonexistent in nature[1]. MMs have an abundance of applications, fabrication schemes, and designs, all of which depend of the particular bandwidth of interest, and desired functionality. *Optical* MMs have been extensively studied, both theoretically and experimentally [2,3,4]. In the visible region of the E-M spectrum, metals such as gold and silver are most commonly used[3]. In the near infrared (NIR) however, heavily doped semiconductors can replace metals as the conductive layer[5]. We can tailor the optical properties of metal oxides in the longer wavelength regime with high levels of extrinsic doping (~$10^{22}$ cm$^{-3}$), without the losses associated with interband transitions in metals.

Hyperbolic Metamaterials (HMMs) create optical anisotropy using either alternating layers with real permittivity having opposite signs, or metallic nanowires suspended in a dielectric media (type I or type II). These geometries create artificial anisotropy, and moreover, an enhanced density of states. Engineering these optical properties have led to the development of many new and exciting areas of study, such as hyperlensing[6,] enhancement of spontaneous quantum emitters[7], selective broadband light absorbers, and so on.

In type I HMM, the dispersion relation goes from an iso-surface (spherical) to ellipsoid (birefringent), and finally a hyperboloid. An E-M wave propagating through anisotropic media is described by the dispersion relation:

$$\omega^2 / c^2 = (k_x^2 + k_y^2)/\varepsilon_\perp + (k_z^2)/\varepsilon_{ll}, \quad (1)$$

where $\varepsilon_\perp$ and $\varepsilon_\parallel$ are the perpendicular and parallel permittivities (relative to direction of E-M wave propagation), $\omega$ and $c$ is the frequency and speed of light, and ($k_x$, $k_y$, $k_z$) are the (x, y, z) components of the incident wave vector respectively. This relationship takes the form of an open hyperboloid when $\varepsilon_\perp \varepsilon_\parallel < 0$. Either $\varepsilon_\perp$ or $\varepsilon_\parallel$ has to be negative in the wavelength range of interest to observe a hyperbolic dispersion transition.

In the NIR region, doped transparent metal oxides have metal-like conductivity, and an associated plasma frequency $(\omega_P)$, that can be described by the Drude theory:

$$\omega_p^2 = ne^2(\varepsilon_o m_0), \quad (2)$$

Where $\varepsilon_o$ is free space permittivity, $e$ is the electron charge, and $n$ is the free carrier density. The plasma frequency is related to the real part of the permittivity through the relation:

$$\varepsilon(\omega) + i\varepsilon(\omega) = \varepsilon_{\text{int}} - (\omega_p^2)/(\omega(\omega + i\Gamma)), \quad (3)$$

Where $\varepsilon_{\text{int}}$ is the permittivity due to interband transitions, and $\Gamma$ is the damping rate. At sufficient carrier concentrations, the real permittivity of the AZO is negative in the NIR and beyond, and satisfies the requirements for the hyperbolic dispersion relation and subsequent MM behavior.

Our previous studies show that sputtered AZO at elevated temperatures has high optical (+80%) transmittance in the visible region, low ambient electrical resistivity (~1 x $10^{-3}$ $\Omega$–cm), and negative optical permittivity in the NIR (cross over wavelength of 1800 nm)[9]. These values were obtained from AZO films with a thickness of 270nm. The AZO films fabricated in this study are much thinner, around 50nm. The resistivity of AZO and ZnO films in this work were measured to be 5 x $10^{-3}$ and 2 x $10^{-2}$ $\Omega$-cm, respectively. The thinner AZO and ZnO films are subjected to higher stress, which leads to the creation of defects. Defects act as trap sites for free charge carriers, resulting in an increase in resistivity and reduction of the plasma frequency. Although the resistivity is higher for the thinner AZO and ZnO films, the carrier concentration was high enough to see a positive – to – negative transition in the real permittivity. Below a thickness of ~ 40 nm the free carrier concentration greatly diminishes, and no positive – to – negative transition in the real permittivity is observed from the visible to NIR. Therefore, a minimum layer thickness of 45-50 nm is required to observe the negative permittivity in the AZO films, which is needed to satisfy the hyperbolic dispersion conditions. Ideally, we want the films to be as thin as possible. As long as the layer thickness is much smaller than the wavelength ranges of interest (t << λ), it is safe to assume homogenization of the ML structure.

In this paper, we explore the utilization of RF – sputtered AZO and ZnO as the constituent layers of an infrared (IR) HMM. HMMs with AZO using different deposition techniques, such as pulsed laser deposition (PLD) and atomic layer deposition (ALD), have been previously explored[4,10]. Their findings show conformal hyperbolic MMs, however the fabrication time for one sample can be over 8 hours. Comparable behavior can be achieved by using a much quicker, cheaper, and robust deposition technique such as RF sputtering.

In the sputtered AZO HMMs, we utilized ZnO as the dielectric, and AZO as the "metal" in the alternating layer by layer configuration (type I). We investigated the optical properties of the constituent AZO and ZnO, and calculated the hyperbolic dispersion transition wavelength using the nonlocal corrected Maxwell-Garnett Effective Medium Approximation (EMA)[10]. The Hyperbolic Behavior is confirmed via modeling of Spectroscopic Ellipsometry data, as well as polarized transmittance measurements.

**Experimental Details**

AZO and ZnO thin films were grown on n-type (001) Si, (001) sapphire, and plain glass by the RF-magnetron sputtering tool (Lab 18 by Kurt J. Lesker). Prior to deposition and substrate loading, a Titanium target was sputtered at 200W for 1 hour to remove any contaminants inside the chamber. The subsequent chamber base pressure dropped an order of magnitude from $10^{-7}$ to $10^{-8}$ Torr overnight. The film substrates were prepared by sonication in acetone, methanol, and isopropanol for 15 minutes each. After, the substrates were dried by $N_2$ gas and immediately transferred into the deposition chamber via a load lock arm. The samples were placed on a parallel sample holder, directly above the targets approximately 70 mm ± 3 mm.

For the AZO film deposition, the power was 150 W, using a ceramic 2 wt%. $Al_2O_3$ doped ZnO target. The ZnO films were grown using the same deposition parameters with a 99.9 wt%. ZnO target installed on another sputtering gun inside the same chamber. The base pressure for all film growths was 2 x $10^{-7}$ Torr. The working pressure during film growth was 5 x $10^{-3}$ Torr with a fixed Ar pressure of 2.5 mT (flow rate 41 sccm) controlled by a mass flow controller. The substrate temperature was fixed at 400°C with a platen motion at 10 rpm to ensure film uniformity across the substrates. The deposition time varied between 5-10 minutes depending on the desired thickness of the individual layers (35-55nm). After deposition, substrate heating was turned off, and the films were left to cool down to ambient temperatures.

The AZO/ZnO multilayers (MLs) were grown by alternating between the AZO and ZnO targets during the sputtering process. Each cycle of AZO and ZnO had the same deposition conditions (time, temperature, RF power, and flow rate). This results in individual film thicknesses of ~50nm ± 2nm for both AZO and ZnO. Several samples were made, having a different number of AZO/ZnO cycles. The data presented in this letter is for samples with 6 cycles of AZO/ZnO (12 total layers), resulting in a total slab thickness around 600nm. The growth rate for both AZO and ZnO at 400°C was approximately 5nm/minute. The resistivity of the films was measured using the 4-probe technique. The optical properties of the constituent films as well as the multilayer slabs were measured via UV-Vis-IR spectroscopy and Variable Angle Spectroscopic Ellipsometry (VASE). The VASE HS 190 ellipsometer from J.A. Woollam company was used, and the data was modeled with WVASE 32 version 3.774 software.

### Results and Discussion

The thickness of the constituent AZO/ZnO films, and AZO/ZnO Multi-Layer (ML) samples were confirmed via cross-sectional SEM and AFM scans. (**Figure 1**) shows a cross-sectional SEM image of the AZO/ZnO ML structure grown on silicon. The higher contrast of the AZO is due to high secondary electron emission from the conductive AZO layers. It is clear from the spectrograph that there are no discontinuities in the deposited layers, nor any adhesion or bowing issues with the substrates. The stark change in contrast outlines clear interfaces between the AZO and ZnO layers. The columnar growth of the ML is apparent from the SEM micrograph showing preferential C-axis (002) orientation. Various studies have confirmed preferential C-axis growth[11,12].

The advantage of using AZO/ZnO layers is they both have the same polycrystalline phases - from a crystallographic view, the materials are identical. This minimizes stress between layers, and reduced the overall optical scattering at the interfaces between films. There is no observable change in the crystallinity as a result of the number of layers deposited – more layers result in higher attenuation of IR, due to the non-zero imaginary permittivity in the longer wavelength range.

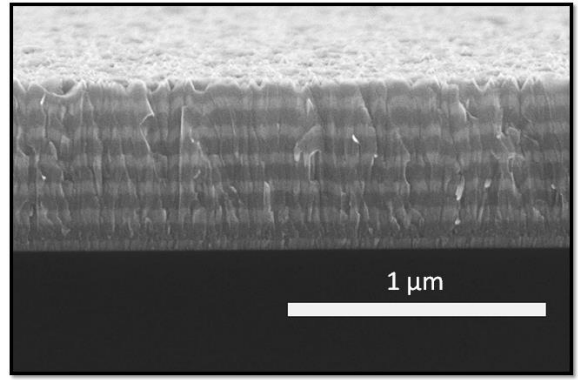

**Figure 1**: Cross-sectional Scanning Electron Micrograph of a 12 layer slab taken at a 4° tilt. The AZO layers are lighter in contrast.

(**Figure 2**) shows UV-Vis-IR transmittance results of the individual ZnO and AZO layers, as well as the multilayer slab. The inset is an optical image of the ML sample grown on glass. The film transmittance was measured with the PerkinElmer Lambda 950, from 350 to 1200nm in 0.5nm increments. The transmittance was measured against a bare glass substrate as a reference to isolate the response of the sputtered films. The Individual ZnO and AZO films are highly transparent, close to 80% in the visible range. Although the multilayer samples are thicker, they are even more transparent in some portions of the visible region, showing transmittance upwards of 90%. This increase in transmittance around 450nm is due to constructive/destructive Fresnel reflections at the AZO/ZnO interfaces and substrate. This shows that there is a distinct change in the optical properties between ZnO and AZO in the visible region. Ellipsometry scans further support this notion, showing a stark contrast in the real part of the permittivity between AZO and ZnO that begins around 450 nm (shown elsewhere).

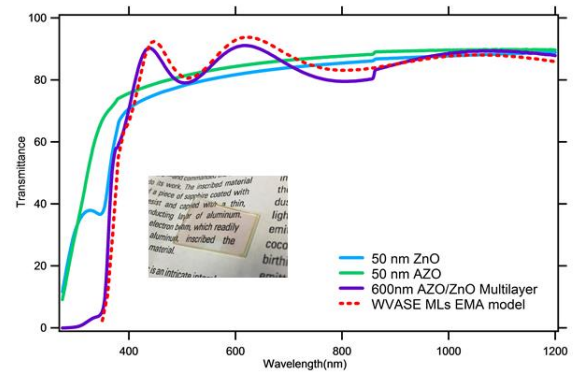

**Figure 2** Transmittance of individual ZnO and AZO films and multilayer stack. The red dotted curve is the simulated transmission using the optical properties extracted from ellipsometry. The inset show a photograph of the AZO/ZnO ML on paper.

Variable Angle Spectroscopic Ellipsometry (VASE) scans of the individual AZO and ZnO films grown on silicon were used to find the real and imaginary permittivity of the films. VASE is one of the most powerful optical characterization tools, capable of measuring the change of intensity ($\psi$) and polarization ($\Delta$) of any material in question - it is especially useful for determining the optical response of complex multilayered samples[13]. They are related through the equation:

$$tan(\psi)e^{i\Delta} = r_p / r_s, \qquad (4)$$

where, $r_p$ and $r_s$ are the Fresnel coefficients for p and s – polarized light.

The AZO and ZnO films, as well as the ML slab, were deposited and immediately taken for ellipsometry measurements to minimize the effects of surface contamination. The samples taken for VASE measurements were deposited on silicon substrates. The measurements ranged from 350-2400nm in 10nm increments, at scan angles of 65,70, and 75° in a clean room environment. The scans resulted in over 500 data points to be used in generating the most accurate model possible.

The physical model used for the measured data was an infinitely thick Si substrate layer (the substrates used were measured and modeled prior to film deposition), followed by the deposited thin film (ZnO, AZO, or MLs), then a surface roughness layer. The dispersion model used for the ZnO was a combination of Lorentzian and Gaussian oscillators. For the AZO, a combination of Drude, Lorentzian, and Gaussian oscillators was used. A total of 5 oscillators were used for both AZO and ZnO. The thickness of the individual films was set to 50 nm, while the dispersion coefficients and surface roughness were varied to best fit the measured data. The Mean Square Error (MSE) between the model and experimental data for individual AZO and ZnO films were below 5.

(**Figure 3A**) shows the Ellipsometry results of the individual films. The ZnO films behave purely as a dielectric throughout the visible and infrared region with extremely low loss. The AZO is dielectric in the visible range, but has a cross over from positive to negative values of permittivity around 2235 nm. Beyond this the AZO exhibits epsilon-near-zero (ENZ) and 'metallic' behavior, satisfying the conditions needed for the type I metamaterial. Although the imaginary permittivity slightly increases (increased optical loss), it is still orders of magnitude lower than that of noble metals which are conventionally used in HMMs.

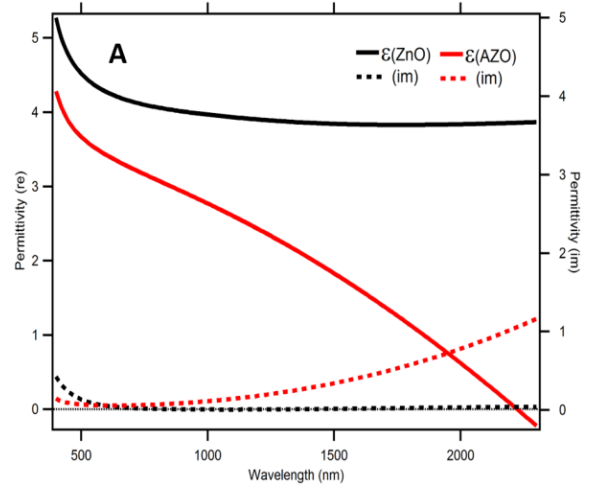

**Figure 3 (A)** Ellipsometry results of individual ZnO and AZO films; both films were 50 nm thick. The ZnO is purely dielectric with low loss from the visible to the infrared. The real permittivity of 50 nm sputtered AZO goes to negative values at 2200 nm.

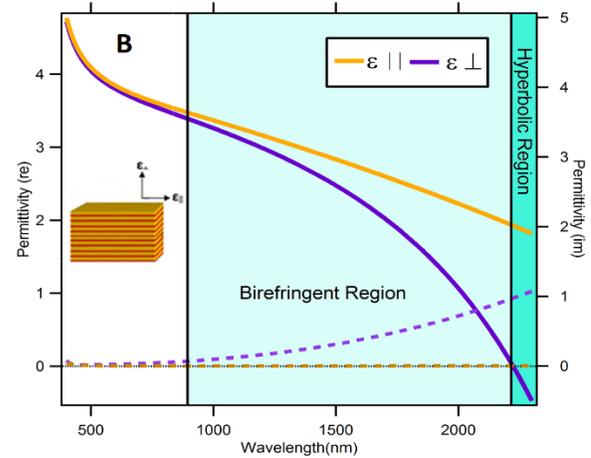

**Figure 3 (B)** The calculated parallel and perpendicular permittivity of the ZnO/AZO ML slab using the nonlocal-corrected EMA. The calculations show the hyperbolic transition occurs around 2225 nm.

The optical properties of the AZO and ZnO films extracted from the VASE scans were used to predict the behavior or the multilayered system using the Effective Medium Approximation (EMA). The predicted optical response was then compared to the measured optical response of the ellipsometry scans. Two different modeling techniques were used to describe the anisotropic optical response of the ML VASE scans. The first model used to fit the VASE data was the Maxwell-Garnett Effective Medium

Approximation (EMA). In the EMA model, the ML sample is considered a homogeneous optical medium, having anisotropic behavior (polarization dependent permittivity). The EMA model enables us to calculate the associated permittivity of a layered material if the thickness, permittivity, and number of layers are known. The first order perpendicular and parallel permittivity ($\varepsilon_\perp$ and $\varepsilon_\parallel$ respectively) can be described as:

$$\varepsilon_\perp = \frac{(t_1+t_2)\varepsilon_1\varepsilon_2}{(t_2\varepsilon_1)+(t_1\varepsilon_2)}, \quad (5)$$

$$\varepsilon_\parallel = \frac{t_1\varepsilon_1 + t_2\varepsilon_2}{t_1+t_2}, \quad (6)$$

Where $t_1$ and $\varepsilon_1$ are the thickness and permittivity of the metallic layer (AZO), and $t_2$ and $\varepsilon_2$ are the thickness and permittivity of the dielectric (ZnO). In order to extract useful information that describes the real response of the system, we have to consider the nonlocal effects between layers. It has been shown that the optical response of multilayers materials is strongly affected by large field oscillations across the system[13]. In order to reconcile the difference, non-local corrections to the ellipsometry/EMA data was implemented. The effective permittivities then become:

$$\varepsilon_\perp = \frac{\varepsilon_{\perp(0)}}{1-\delta_\perp(k,\omega)}, \quad (7)$$

$$\varepsilon_\parallel = \frac{\varepsilon_{\parallel(0)}}{1-\delta_\parallel(k,\omega)}, \quad (8)$$

the nonlocal corrections take the form of:

$$\delta_\perp = \frac{(t_1^2 t_2^2 (\varepsilon_1-\varepsilon_2)^2 \varepsilon_\parallel^{(0)2}))}{(12(t_1+t_2)^2 \varepsilon_1^2 \varepsilon_2^2)} [\varepsilon_\parallel^{(0)} \frac{\omega^2}{c^2} - \frac{k_\perp^2(\varepsilon_1+\varepsilon_2)^2}{(\varepsilon_\parallel^{(0)2})}], \quad (9)$$

$$\delta_\parallel = \frac{(t_1^2 t_2^2 (\varepsilon_1-\varepsilon_2)^2)}{12(t_1+t_2)^2 \varepsilon_\parallel^{(0)}} \frac{\omega^2}{c^2}. \quad (10)$$

The results of the non-local corrected EMA calculations are shown in (**Figure 3B**). T-E polarized light shows purely dielectric behavior from UV to NIR with very low loss. T-M polarized light shows the hyperbolic transition occurs at 2.3 µm. The imaginary permittivity increases significantly after the transition, causing significant attenuation of light above that wavelength. This is detrimental for waveguide/ optical sensing designs. However, the increase in light absorption may be essential for infrared beam steering and broadband absorption applications.

The second data modeling method uses a uniaxial material to describe the optical response of the ML sample with generalized ellipsometry scans. Anisotropic materials have non-zero off diagonal components of the *Jones Matrix*, which is a 2 x 2 representation of an optical system that changes the polarization state of light:

$$\begin{pmatrix} \hat{E}_p \\ \hat{E}_s \end{pmatrix} = \begin{pmatrix} r_{ss} & 0 \\ 0 & r_{pp} \end{pmatrix} \begin{pmatrix} \hat{E}_p^{in} \\ \hat{E}_s^{in} \end{pmatrix}, \quad (11)$$

Mixing of p and s – polarized light can occur when propagating through anisotropic media. This leads to non-zero values of the off diagonal elements of the Jones matrix. The off diagonal elements of the Jones matrix need to be considered when describing the optical response of an anisotropic medium (MLs).:

$$\begin{pmatrix} \hat{E}_p \\ \hat{E}_s \end{pmatrix} = \begin{pmatrix} r_{ss} & r_{sp} \\ r_{ps} & r_{pp} \end{pmatrix} \begin{pmatrix} \hat{E}_p^{in} \\ \hat{E}_s^{in} \end{pmatrix}, \quad (12)$$

Generalized Ellipsometry measures the off-diagonal elements through careful selection of the parameters during setup (fixed polarization at more than one angle). The off diagonal elements are related through the equations, $tan(\psi_{ps})e^{(i\Delta_{ps})} = r_{ps}/r_{pp}$, and $tan(\psi_{sp})e^{(i\Delta sp)} = r_{sp}/r_{ss}$. These coefficients are needed to determine the anisotropic response of the ML samples. The ML samples were measured in anisotropic mode, From 350-2300 nm, in 10 nm increments, with 3 scan angles of 65, 70, and 75°. Both modeling techniques offer a unique and independent approach to extracting the optical parameters of interest.

Fitting the generalized ellipsometry data was done by using both a uniaxial material model, and an EMA material model, with initial dispersion parameters extracted for the EMA calculation[14]. The combination of these models enables the extraction of directionally dependent permittivity values of the ML slabs. These models use the optical properties of the constituent AZO and ZnO films, the filling fraction of the constituent materials (assumed to be 0.5 when AZO and ZnO have equal thickness), the Euler angles, depolarization factor, and finally the total slab thickness. The thickness was fixed at 630 nm (confirmed from SEM cross sections). The only free fitting parameters were the filling fraction, surface roughness, and depolarization. A point-by-point fitting of the VASE data was done, while Kramers – Kronig constraints were enforced throughout the oscillator fitting of the permittivity dispersion.

Using the optical properties extrapolated from the Ellipsometry data we were able simulate the transmission of the ML slabs (red curve in **Figure 2**). From the uniaxial material model, the hyperbolic transition takes place at 2245nm, just 20 nm off from the EMA-calculated value (2225nm). The small discrepancy is most likely due the deviation of the thickness of the deposited films as well as the surface roughness. From the initial EMA calculations, the thickness of the AZO and ZnO layers was set equal (filling factor of 0.5). The corrected ellipsometry results suggest that the ML slab was 47% ZnO and 53% AZO. This slight deviation is attributed to the small difference in the film thickness deposited during sputtering.

The EMA model had higher Mean Square Error (MSE) due to the fact that the depolarization factor is fixed during the fitting process, unlike the uniaxial material model where mixing of S and P – polarized light is measured and included in the modeling. In both fitting techniques, there was relatively good agreement between the calculated and fitting experimental data. **Table I** shows the comparative fitting results of the measured generalized ellipsometry data, using the nonlocal corrected EMA calculations, EMA and uniaxial material models. The quality of the model fitting depended on the degree of depolarization and number of layers deposited (12 total layers in this case). The TE polarized data was a poor fit to our physical model, regardless of the number of layers or dispersion coefficients. The TM polarized light fit very well, with MSE under 9 for all samples measured. Therefore, a most of the error between modeled and measured data is due to the off-axis polarized light. Fitting the off axis polarization data remains a challenge to the validity of the EMA model as it applies to complex multilayered systems.

**Table I** shows the calculated parallel/perpendicular permittivity values using EMA, as compared to VASE data modeled with EMA and uniaxial material model.

| ML model | HMM transition | $\varepsilon_\parallel$ @ $1.5\ \mu m$ | $\varepsilon_\perp$ @ $1.5\ \mu m$ | MSE |
|---|---|---|---|---|
| calculated | 2225 nm | 2.9+$i$0.007 | 2.1+$i$0.27 | n/a |
| EMA | 2295 nm | 3.2+$i$0.005 | 2.003+$i$0.3 | 19.403 |
| Uniaxial | 2245 nm | 3.39+$i$0.004 | 1.95+$i$0.312 | 14.506 |

To demonstrate the hyperbolic behavior of the AZO/ZnO multilayers we performed the beam blocking transmittance measurement[9,15]. The experimental setup is shown in the inset of (**Figure 4**). Exactly half of the photodetector is blocked at normal incidence relative to the ML sample. the VASE alignment pin was used to ensure half of the photodetector was blocked; A razor blade was placed in the optical path of the system, in between the source and detector, after the sample stage. The razor was mounted on a translation stage perpendicular to the optical axis of the system, and translated such that all the light reflected off of the razor and reaching the alignment pin was in the first and fourth quadrant of the pin photodetectors. This ensured that the blade was impeding exactly half of the light reaching the VASE detector. The spot size of the incident beam was approximately 1 mm. The beam – blocking blade was 150 mm away from the detector.

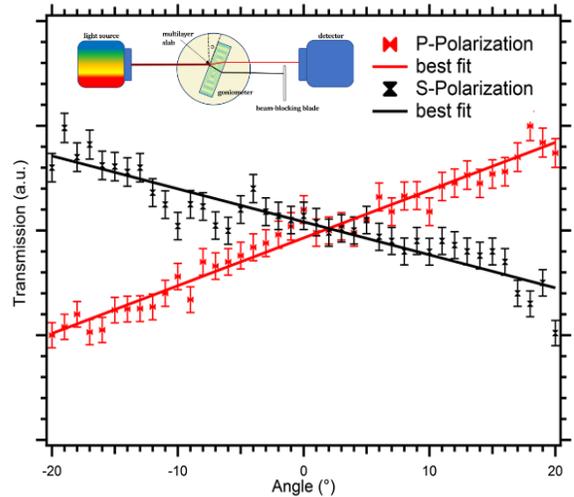

**Figure 4** beam – blocking transmittance measurements. S and P - polarized light have transmission trends of opposite signs, indicating negative refraction (dotted).

The relative transmittance of the ML samples was measured as a function of incidence angle from -20° to 20° in 1° increments. The wavelength of the incident light was fixed at 2.4 μm, which is beyond the wavelength that our ML slab undergoes a hyperbolic transition (around 2.2 μm). In isotropic materials, the S and P – polarized transmittance angles are the same, independent of the incidence angle. Anisotropic materials transmit S and P – polarized light at different angles, but the overall trend is the same. For our ML slab, the opposite trends were observed**. (Figure 4)** shows the P – polarized light *increasing* in relative transmittance, indicating negative refraction. The S – polarized light decreases in relative intensity, typical behavior of isotropic dielectric materials. This verifies that our sample is indeed anisotropic, capable of Type I hyperbolic behavior.

**Conclusion**

In conclusion, ZnO/AZO hyperbolic metamaterials were fabricated using the RF sputtering deposition technique. Optical characterizations show that the individual AZO and ZnO layers do indeed vary in optical properties, with AZO having metallic nature in the IR, satisfying the condition for a type I hyperbolic metamaterial. Spectroscopic Ellipsometry and angular transmission results confirm that our sample has a transverse positive hyperbolic transition in the infrared region. Future work includes optimization and post processing of the ML structure, and dynamic tunability of the effective permittivity. To our knowledge, this is the first HMM grown using RF sputtering as a deposition process. This work adds an additional paradigm of fabrication in the field of infrared plasmonics, metamaterials, and complex light manipulation on the nano scale. The main draw back here is that there is a compromise between precision of deposition (thickness and surface roughness) and throughput. ALD and PLD allow angstrom level precision at the expense of time and cost. If this level of precision is not necessary, RF sputtering promises to be a suitable alternative for robust optical applications.

**Acknowledgments**

KS conceived the idea and wrote the main text of the paper. KS, RM, and CW conducted the experiments. MB and AP provided academic and technical guidance and manuscript review. This work is supported by the NSF-CREST Grant number HRD 1036494 and HRD 1547771.

**Corresponding author**

Correspondence to: Dr. Kevin C. Santiago (ksantia1@tnstate.edu)

[1] X. Zhang, "Metamaterials: a new frontier of science and technology," *Chem. Soc. Rev .,*2011,40,2494–250

[2] Y. H. R. Mittra, FDTD Modeling of Metamaterials, Boston: Artech House, 2009.

[3] N.M. Litchinitser, "Metamaterials: transforming theory into reality," *Opt. Soc. Am. B,* vol. 26, no. 12, pp. 161-169.

[4] T.Riley, "High Quality, Ultraconformal Aluminum-Doped Zinc Oxide Nanoplasmonic and Hyperbolic Metamaterials," *Material Views,* vol. 12, no. 7, pp. 892-901, 2016.

[5] A.K.Pradhan, "Extreme Tunability in Aluminum Doped Zinc Oxide plasmonic materials for near IR applications," *Nature Scientific Reports,* 2014. doi:10.1038/srep06415

[6] Z. haowei Liu, "Far-Field Optical Hyperlens Magnifying Sub-Diffraction-Limited Objects," *Science,* vol. 315, no. 5819, p. 1686, 2007.

[7] L. Ferrari, "Enhanced spontaneous emission inside hyperbolic Metamaterials," *Optics Express,* vol. 22, no. 4, pp. 4301-4306, 2014.

[8] V. Sreekanth, "A multiband perfect absorber based on hyperbolic metamaterials," *Nature: Scientific Reports,* pp. 1-8, 2016.

[9] H. Dondapati, "Influence of growth temperature on electrical, optical, and plasmonic properties of Aluminum: Zinc oxide films grown by radio frequency magnetron sputtering," *Journal of Applied Physics*, 2013 https://doi.org/10.1063/1.4824751.

[10] G. V. Naik, "Demonstration of Al:ZnO as a plasmonic component for NIR Metamaterials," *PNAS,* pp. 1-5, 2011. doi:10.1073/pnas.1121517109

[10] O. Kidwai, "Effective-medium approach to planar multilayer hyperbolic metamaterials: Strengths and limitations," *Phys. Rev. A* 85, 053842

[11] C. Tseng, "Effects of sputtering pressure and Al buffer layer thickness on properties of AZO films grown by RF Magnetron Sputtering," *Elsevier,* pp. 263-267, 2010. doi:10.1016/j.vacuum.2010.06.006

[12] C. B. Park, "The Electrical and Optical Properties of Al-Doped ZnO films in Ar:H2 Radio Frequency Magnetron Sputtering system," *transactions on el. and opt.,* vol. 11, no. 2, pp. 81-84, 2010. http://dx.doi.org/10.4313/TEEM.2010.11.2.081

[13] A. Podolskiy, "Nonlocal effects in effective-medium response of nanolayered metamaterials," *App. Phys. Letters* 90 (2007) https://doi.org/10.1063/1.2737935

[14] P. Kelly, "Designing optical metamaterial with hyperbolic dispersion based on an Al:ZnO/ZnO nano-layered structure using the atomic layer deposition technique," *Applied Optics,* vol. 55, no. 11, pp. 2993-2997, 2016. https://doi.org/10.1364/AO.55.002993

[15] T. Tumkur, "Permittivity evaluation of multilayered hyperbolic metamaterials: Ellipsometry vs. reflectometry," *Applied Phys letters,* no. 117, 2015. https://doi.org/10.1063/1.4914524

[16] Y. Zhong, "Review of Mid Infrared Plasmonic Materials," J. Nanophoton. 9(1) 093791 doi:10.1117/1.JNP.9.093791

[17] M. Noginov, "Bulk photonic metamaterial with hyperbolic dispersion". Optics/International Quantum Electronics Conference, OSA Technical Digest (CD) (Optical Society of America, 2009), paper JWC2. https://doi.org/10.1364/CLEO.2009.JWC2

[18] E. Lim, "Review of Silicon Photonics Foundry Efforts," *IEEE Journal of Selected Topics,* vol. 20, no. 4, p. 830012, 2014.

[19] W. Dang, "Deposition and characterization of sputtered ZnO films," *Science Direct; Superlattices and Microsctructures,* pp. 90-93, 2007. doi:10.1016/j.spmi.2007.04.081

[20] P. Drude, "Zur Elektronentheorie der Metalle," vol. 1, p. 566, 1900. Doi:10.1002/andp.19003060312

[21] G. Garcia, " Near-Infrared Spectrally Selective Plasmonic electrochromic thin films," *Advanced Optical Materials,* pp. 215-220, 2013. DOI: 10.1002/adom.201200051

[22] J. Kim, "Plasmonic Resonances in Nanostructured Transparent Conducting Oxide Films," *IEEE Journal of Selected Topics,* vol. 19, no. 3, 2013.

[23] M. Sriram, "Single Nanoparticle Plasmonic Sensors," *Open Access: Sensors,* pp. 25774-25792, 2015. doi:10.3390/s151025774

[24] S.A.Maier, Plasmonics: Fundamentals and applications, New York: Springer, 2007.

[25] S. Schlcker, "Surface Enhances Raman Spectroscopy: Concepts and Chemical Applications," *Angewandte Reviews,* pp. 4756-4795, 2014. DOI: 10.1002/anie.201205748

[26] M. Shalaev, "Optical Negative Index Metamaterials," *Nature Photonics,* vol. 1, pp. 41-48, 2007. doi:10.1038/nphoton.2006.49

[27] M. I. Stockman, "Nanoplasmonics: past, present, and a gilmpse into future," *Optical Society of America,* vol. 19, no. 22, pp. 22029-22126, 2011. doi.org/10.1364/OE.19.022029

[28] P. Kelly, "Designing optical metamaterial with hyperbolic dispersion based on an Al:ZnO/ZnO nano-layered structure using the atomic layer deposition technique," *Applied Optics,* p. 1559, 2016.

[29] L. Brekhovskih, Waves in Layered Media: Second Edition, New York: Academic Press, 1980.

[30] H. G. T. Eugene A. Irene, Handbook of Ellipsometry, Norwich, NY: William Andrew Publishing, 2005.